\documentstyle[preprint,aps,epsf,floats]{revtex}

\begin{document}

\tightenlines

\font\fortssbx=cmssbx10 scaled \magstep2
\hbox to \hsize{
\hfill
\vtop{\hbox{\bf MADPH-00-1160}
      \hbox{\bf Fermilab-PUB 00/049-T}
      \hbox{\bf AMES-HET 00-01}
      \hbox{March 2000}}}
\vspace{.5in}

\begin{center}
{\large\bf Neutrino Oscillations at an Entry-Level\\
Neutrino Factory and Beyond}\\[6mm]
V. Barger$^1$, S. Geer$^2$, R. Raja$^2$, and K. Whisnant$^3$\\[3mm]
\it
$^1$Department of Physics, University of Wisconsin,
Madison, WI 53706, USA\\
$^2$Fermi National Accelerator Laboratory, P.O. Box 500,
Batavia, IL 60510, USA\\
$^3$Department of Physics and Astronomy, Iowa State University,
Ames, IA 50011, USA

\end{center}

\begin{abstract}
We consider the parameters of an entry-level neutrino factory
designed to make the first observation of $\nu_e\to\nu_\mu$
oscillations, measure the corresponding amplitude $\sin^22\theta_{13}$,
and determine the sign of the atmospheric-scale $\delta m_{32}^2$ via matter
effects. A 50~kt detector, a stored muon energy $E_\mu \geq 20$~GeV 
and $10^{19}$ muon decays would enable these goals to be met provided 
$\sin^22\theta_{13}>0.01$. The determination of the sign of $\delta m_{32}^2$ 
also requires a baseline $L \ge 2000$~km. An upgraded neutrino factory   
with O($10^{20}$) decays would enable the first observation of 
$\nu_e\to\nu_\tau$ oscillations. With O($10^{21}$) decays
the effects of a large CP-phase could be measured in the case of 
the large angle matter oscillation solution to the solar neutrino anomaly. 
Our analysis includes a family of three-neutrino models that can account 
for the atmospheric and solar neutrino oscillation indications.
\end{abstract}

\thispagestyle{empty}
\newpage

\section{Introduction}

The observation of a deficit
of muon-neutrinos in atmospheric neutrino experiments\cite{atmos1,atmos2} 
has paved the way for a new generation of experiments studying neutrino 
masses and mixing. The neutrino sector offers exceptional opportunities for 
studying some of the most fundamental issues in particle physics, 
such as the origin of masses and CP violation. 
A major advantage over the quark sector is 
that neutrino phenomena are free of the complications of strong interactions. 
A comprehensive knowledge 
of the neutrino mixing matrix may yield clues to the old puzzle 
of why there is more than one lepton family.

The SuperKamiokande (SuperK) collaboration\cite{atmos1} has found that
the atmospheric neutrino deficit is dependent on $L/E_\nu$, with greater 
suppression of the
$\nu_\mu$ flux with increasing $L/E_\nu$. Moreover, the electron-neutrino rate
is $L/E_\nu$ independent and consistent with the calculated
$\nu_e$ flux. The natural interpretation of the atmospheric data is in terms of
$\nu_\mu\to\nu_\tau$ oscillations, with maximal or near-maximal mixing and a
neutrino mass-squared difference $\delta m^2_{\rm atm} \simeq
3\times10^{-3}\rm\,eV^2$.
This is supported by SuperK measurements of the zenith angle
dependence, which due to matter effects is different for
$\nu_\mu\to\nu_\tau$ and $\nu_\mu\to\nu_s$, and by $\pi^0$
production in neutral current events, which taken together rule out
$\nu_\mu\to\nu_s$ at 99\%~C.L~\cite{kajita99}.

There are other indications of neutrino oscillation phenomena at $\delta m^2$
values distinct from the $\delta m^2_{\rm atm}$ scale. A long-standing puzzle
is the observed deficits of solar neutrinos\cite{solar1} compared to the flux
predictions of the Standard Solar Model\cite{solar2}. There are four
regions\cite{solardata} of oscillation parameters $\delta m^2_{\rm solar},\
\sin^22\theta_{\rm solar}$ that can accommodate the present solar data. Three
of the solutions involve resonance
enhancements\cite{wolf-etal,bppw80,langacker,mikheyev,parke-etal} due to the
coherent scattering of $\nu_e$ from the dense solar medium. The Large Angle
Matter (LAM) solution has $\delta m^2_{\rm solar}\sim 3\times10^{-5}\rm\,eV^2$,
the Small Angle Matter (SAM) solution has $\delta m^2_{\rm solar} \sim
5\times10^{-6}\rm\,eV^2$, and the Long Oscillation Wavelength (LOW) solution
has $\delta m^2_{\rm solar} \sim 10^{-7}\rm\,eV^2$ and large mixing. Vacuum
Oscillation (VO) solutions\cite{bpw81,vlw} have $\delta m^2 \sim
10^{-10}\rm\,eV^2$ and large mixing.  The Sudbury Neutrino Observatory (SNO)
experiment\cite{sno} in progress may be able to exclude some of these solar
solutions.

In addition there is some evidence for $\nu_\mu\to\nu_e$ and
$\bar\nu_\mu\to\bar\nu_e$ oscillations from an accelerator experiment
(LSND) at Los Alamos\cite{lsnd}. The observed event rates correspond to
$\sin^22\theta_{\rm LSND}\sim10^{-2}$ with $\delta m^2_{\rm LSND}\sim
1\rm\,eV^2$; a sizeable range of $\delta m^2$ values above 0.1~eV$^2$ is
actually allowed. The mini-BooNE experiment at Fermilab\cite{miniboone},
scheduled to start collecting data in December
2001 with first results expected by 2003, will determine whether the LSND
effect is real.

In the next phase of neutrino oscillation studies long-baseline experiments 
are expected to
confirm the $\nu_\mu\to\nu_\mu$ disappearance oscillations at the $\delta
m^2_{\rm atm}$ scale. The K2K experiment\cite{k2k} from KEK to SuperK,
with a baseline of $L\simeq 250$~km and a mean neutrino energy of $\left<
E_\nu\right>
\sim 1.4$~GeV is underway. The MINOS experiment from Fermilab to
Soudan\cite{minos}, with a longer baseline $L\simeq 730$~km and higher mean
energies $\left< E_\nu \right> = 3$, 6 or 12~GeV, is under construction and the
ICANOE\cite{icanoe} and OPERA\cite{opera} experiments, with $\left<E_\nu\right> = 17$~GeV and baselines $L\simeq 730$~km 
from CERN to Gran Sasso, have been proposed.
The various experiments with dominant $\nu_\mu $ and $\bar\nu_\mu$ beams will
securely establish the oscillation phenomena and may measure $\delta m^2_{\rm
atm}$ to a precision of order 10\%\cite{thesis}.
Experiments with higher mean neutrino energies should be able
to observe $\tau$ production.

Further exploration of the neutrino mixing and mass-squared parameter space
will require higher intensity neutrino beams and $\nu_e,\ \bar\nu_e$ beams
along with $\nu_\mu,\ \bar\nu_\mu$. To provide these neutrino beams, 
muon storage rings have been proposed~\cite{geerconf,geer} in which the muons 
decay in a long straight neutrino beam-forming section, and the muons are 
produced by a muon-collider type muon source~\cite{mucoll}. 
These ``neutrino factories" are now under serious
consideration\cite{geerconf,geer,mucoll,derujula,camp,nu-factory,BGW,BGRW,cervera,shrock}. The
resulting neutrino beams would be sufficiently intense to produce thousands of
oscillation events in a reasonably sized detector (10--50~kt) at distances up
to the Earth's diameter\cite{geer}. Some initial studies have been made of the
physics capabilities of such
machines\cite{derujula,camp,nu-factory,BGW,BGRW,cervera,shrock} as a 
function of the stored muon energy $E_\mu$, baseline $L$, and
intensity $I$. The focus
of the present paper is how to best choose $E_\mu$, $L$ and $I$ 
to maximize the physics output at an 
entry-level neutrino factory (hereafter referred to as ENuF) and beyond. 

The present work expands on previous studies in several ways. First, 
we consider the {\it minimal} $E_\mu$ and $I$ needed to accomplish our 
physics goals. 
Second, we consider the consequences of a number of model scenarios that can
accommodate the atmospheric and solar oscillation
indications.  Third, we investigate possibilities for measuring CP-violating
phases\cite{derujula,cervera,cabbibo,bpw80,pakvasa,barger-CP,CP}. Finally, we 
investigate possibilities for observing $\nu_e \rightarrow \nu_\tau$ 
oscillations.

\section{Theoretical overview}

\subsection{Oscillation Formalism}

The neutrino flavor eigenstates $\nu_\alpha$ are related to the mass
eigenstates $\nu_j$ in vacuum by a unitary matrix $U$,
\begin{equation}
\left| \nu_\alpha \right> = \sum_j U_{\alpha j} \left| \nu_j \right>\,.
\end{equation}
The effect of matter on $\nu_e$
beams~\cite{bppw80,langacker,mikheyev,parke-etal,shrock,bernstein,pantaleone,lipari} has
important consequences for long-baseline experiments.
The propagation through matter is described by the evolution equation
\begin{equation}
i{d\left|\nu_\alpha\right>\over dx} = \sum_\beta \sum_{j\neq1} {1\over 2E_\nu}
\left[ \delta m_{j1}^2 U_{\alpha j} U^*_{\beta j} + A \delta _{\alpha e}
\delta_{\beta e} \right] \left|\nu_\beta\right> \,,
\label{eq:prop}
\end{equation}
where $x=ct$ and $A/(2E_\nu)$ is the amplitude for coherent scattering of
$\nu_e$ on electrons, and
\begin{equation}
A = 2\sqrt2\, G_F\, Y_e\rho E_\nu = 1.52 \times 10^{-4}{\rm\,eV^2}\, Y_e\,
\rho\,({\rm g/cm^3}) E\,({\rm GeV}) \,.
\label{eq:A}
\end{equation}
Here $Y_e(x)$ is the electron fraction and $\rho(x)$ is the matter
density. The neutrino oscillation probabilities are then
$P(\nu_\alpha\to\nu_\beta) = |\left< \nu_\beta(x=L) \mid \nu_\alpha(x=0)
\right>|^2$.

We solve this evolution equation numerically taking into account the
$x$-dependence of the density using the Preliminary Reference Earth
model\cite{PREM}. We have found that for $L$ less than about 3000~km (in
which the entire neutrino path is in the upper mantle and the density
is approximately constant), the results of the exact propagation and
those obtained assuming constant density agree to within a few
percent. However, for larger $L$ (where the neutrino path partially
traverses the lower mantle and the density is no longer nearly constant)
the assumption of constant density is no longer valid. For example, for
$L=7332$~km (the Fermilab to Gran Sasso distance), event rate
predictions assuming constant density can be wrong by as much as 40\%.
We always use the numerical solution of Eq.~(\ref{eq:prop}) in our
calculations.

For three neutrinos (with $\alpha = e,\mu,\tau$ and $j=1,2,3$) the
Maki-Nakagawa-Sakata (MNS)\cite{mns} mixing matrix will be parameterized by
\begin{equation}
U
= \left( \begin{array}{ccc}
  c_{13} c_{12}       & c_{13} s_{12}  & s_{13} e^{-i\delta} \\
- c_{23} s_{12} - s_{13} s_{23} c_{12} e^{i\delta}
& c_{23} c_{12} - s_{13} s_{23} s_{12} e^{i\delta}
& c_{13} s_{23} \\
    s_{23} s_{12} - s_{13} c_{23} c_{12} e^{i\delta}
& - s_{23} c_{12} - s_{13} c_{23} s_{12} e^{i\delta}
& c_{13} c_{23} \\
\end{array} \right) \,,
\label{eq:MNS}
\end{equation}
where $c_{jk} \equiv \cos\theta_{jk}$, $s_{jk} \equiv \sin\theta_{jk}$, and
$\delta$ is the non-conserving phase. Two additional diagonal phases are
present in $U$ for Majorana neutrinos, but these do not affect
oscillation probabilities.

\subsection{Measuring $\theta_{13}$  and the Sign of $\delta
\lowercase{m}_{32}^2$}

The oscillation channels $\nu_e\to \nu_\mu$ and
$\bar\nu_e\to\bar\nu_\mu$ can be explored for the first time at a
neutrino factory. In addition to a first observation of these
transitions, the mixing angle $\theta_{13}$ can be measured, the sign of
$\delta m_{\rm atm}^2$ can be determined from matter effects, and the CP
phase $\delta$ could be measured or bounded. With this information,
models of oscillation phenomena can be tested and discriminated.

The charged current (CC) interactions resulting from $\nu_e\to\nu_\mu$ and
$\bar\nu_e\to\bar\nu_\mu$ oscillations produce ``wrong-sign" muons (muons of
opposite charge from the neutrinos in the beam). In the leading oscillation
approximation the probability for $\nu_e\to\nu_\mu$ in 3-neutrino
oscillations through matter of constant density
is~\cite{bppw80,bernstein,pantaleone}
\begin{equation}
P(\nu_e\to\nu_\mu) = s_{23}^2 \sin^22\theta_{13}^m \sin^2\Delta_{32}^m\,,
\label{eq:Pemu}
\end{equation}
where
\begin{equation}
\sin^22\theta_{13}^m =
{ \sin^22\theta_{13}\over \left( {A\over\delta m^2_{32}} -
\cos 2\theta_{13} \right)^2 + \sin^22\theta_{13} }
\label{eq:amp13m}
\end{equation}
and
\begin{equation}
\Delta_{32}^m = { 1.27 \delta m_{32}^2 \,({\rm eV^2})\,L\,({\rm km)} \over
E_\nu\,\rm(GeV) }\; \sqrt{ \left( {A\over\delta m^2_{32}} - \cos2\theta_{13}
\right)^2 + \sin^22\theta_{13}} \,.
\label{eq:delta32m}
\end{equation}
Here $A$ is the matter amplitude of Eq.~(\ref{eq:A}). Thus even with matter
effects the $\nu_e\to\nu_\mu$ probability is approximately proportional to
$\sin^22\theta_{13}$. The experimental sensitivity of the $\nu_e\to\nu_\mu$
measurements therefore changes almost linearly with $\sin^22\theta_{13}$.

For $\bar\nu_e\to\bar\nu_\mu$ oscillations, the sign of $A$ is reversed
in Eqs.~(\ref{eq:amp13m}) and (\ref{eq:delta32m}). For
$\sin^22\theta_{13} \ll 1$ and $A \sim \delta m^2_{32} > 0$, $P(\nu_e\to\nu_\mu)$ is enhanced  and
$P(\bar\nu_e\to\bar\nu_\mu)$ is suppressed  by matter
effects; the converse is the case for $-A\sim \delta m_{32}^2 < 0$. Thus a comparison of the $\nu_e\to\nu_\mu$ and
$\bar\nu_e\to\bar\nu_\mu$ CC rates gives information on the sign of
$\delta m^2_{32}$.

\subsection{$\nu_e\to\nu_\tau$ oscillations}

The availability of $\nu_e$ and $\bar\nu_e$ beams from a neutrino factory
would allow a search for $\nu_e\to\nu_\tau$ oscillations.
In the leading oscillation
approximation the probability for $\nu_e\to\nu_\tau$ 
oscillations through matter of constant density
is~\cite{bppw80,bernstein,pantaleone}
\begin{equation}
P(\nu_e\to\nu_\tau) = c_{23}^2 \sin^22\theta_{13}^m \sin^2\Delta_{32}^m\,,
\label{eq:Petau}
\end{equation}
Thus the $\nu_e\to\nu_\tau$ probability in matter is also approximately
proportional to $\sin^22\theta_{13}$.

\subsection{$CP$ Violation}

In vacuum, $CP$ violation in the lepton sector can be explored by
comparing oscillation probabilities involving neutrinos with the
corresponding  probabilities for oscillations involving antineutrinos.
For three-neutrino oscillations in a vacuum, the probability difference
is
\begin{equation}
P(\nu_\alpha\to\nu_\beta) - P(\bar\nu_\alpha\to\bar\nu_\beta) =
\mp 4 J (\sin2\Delta_{32}+\sin2\Delta_{21}+\sin2\Delta_{13}) \,,
\label{eq:deltaP}
\end{equation}
where $\Delta_{jk} \equiv 1.27 \delta m^2_{jk} ({\rm eV}^2) L ({\rm km})
/ E_\nu ({\rm GeV})$ and $J$ is the $CP$-violating invariant
\cite{keung,jarlskog}, which can be defined as
$J = Im\{ U_{e2} U_{e3}^* U_{\mu 2}^* U_{\mu 3}\}$. The minus (plus)
sign in Eq.~(\ref{eq:deltaP}) is used when $\alpha$ and $\beta$ are in
cyclic (anticyclic) order, where cyclic order is defined as $e\mu\tau$.
For the mixing matrix in Eq.~(\ref{eq:MNS}),
\begin{equation}
J = {1\over8} \sin2\theta_{12} \sin2\theta_{13} \sin2\theta_{23}
\cos\theta_{13} \sin\delta \,.
\label{eq:J}
\end{equation}
Thus even for $\delta = \pm90^\circ$, $J$ will be small when the $\theta_{12}\,, \theta_{13}$ mixing
angles are small.

For $|\delta m^2_{32}| \gg |\delta m^2_{21}|$, the $CP$-violating
probability difference for $\nu_e\to\nu_\mu$ is given approximately by
\begin{equation}
P(\nu_e\to\nu_\mu) - P(\bar\nu_e\to\bar\nu_\mu) \simeq
-4 J \sin \left( {2.54 \delta m^2_{21}({\rm eV}^2) L({\rm km})\over
E_\nu({\rm GeV})} \right) \,,
\label{eq:deltaP2}
\end{equation}
It is evident from Eq.~(\ref{eq:deltaP2}) that $CP$ violation is only
appreciable in  vacuum when the sub-leading oscillations (in this case
oscillations due to $\delta m^2_{21}$) begin to develop\cite{bpw80}.
The same qualitative results are expected when neutrinos propagate
through matter, although the oscillation probabilities are changed.

\section{Family of Scenarios}

With three neutrinos, there are only two distinct $\delta m^2$ values. The
evidence for atmospheric, solar and accelerator neutrino oscillations 
at three different
$\delta m^2$ scales cannot be simultaneously accommodated in a three-neutrino
framework.
Here we set the accelerator evidence aside and use three-neutrino oscillations
to explain atmospheric and solar data; the oscillation scale of the
accelerator data is more relevant to short-baseline experiments.

\begin{table}[t]
\caption{Representative neutrino oscillation scenarios.\label{tab:scenarios}}
\begin{tabular}{lccccccc}
Scenario& $|\delta m_{32}^2|$& $|\delta m_{21}^2|$& $\sin^22\theta_{23}$&
$\sin^22\theta_{12}$& $\sin^22\theta_{13}$ & $J/\sin\delta$\\
\hline
 & (atmos)& (solar)& (atmos)& (solar)& \\[-1ex]
1) LAM& $3.5\times10^{-3}$& $5\times10^{-5}$& 1& 0.8& 0.04 & 0.022\\
2) SAM& $3.5\times10^{-3}$& $10^{-5}$& 1& 0.01& 0.04 & 0.0025\\
3) LOW& $3.5\times10^{-3}$& $10^{-7}$& 1& 0.9& 0.04 & 0.024\\
4) BIMAX& $3.5\times10^{-3}$& $5\times10^{-5}$& 1& 1& 0 & 0
\end{tabular}
\end{table}

A family of representative scenarios in Table~\ref{tab:scenarios},  defined in
the ongoing Fermilab long-baseline workshop study\cite{fermi-lb}, will be
adopted for our subsequent analysis; the central value $|\delta m_{32}^2| =
3.5\times10^{-3}\rm\,eV^2$ is based on the published SuperK data\cite{atmos1}.
With further data accumulation, a slightly lower central value of
$2.8\times10^{-3}\rm\,eV^2$ is indicated\cite{kajita99}, with a
90\%~C.L. range of $2-5\times10^{-3}$~eV$^2$. The forms of the mixing
matrix $U$ in these scenarios are
\begin{eqnarray}
U({\rm LAM})
&=& \left( \begin{array}{ccc}
0.846 & 0.523 & 0.101 e^{-i\delta} \\
-0.372 - 0.060 e^{i\delta} & 0.602 - 0.037 e^{i\delta} & 0.704 \\
0.372 - 0.060 e^{i\delta} & -0.602 - 0.037 e^{i\delta} & 0.704 \\
\end{array} \right) \,,
\label{eq:U1}\\
U({\rm SAM})
&=& \left( \begin{array}{ccc}
0.994 & 0.050 & 0.101 e^{-i\delta} \\
-0.035 - 0.071 e^{i\delta} & 0.706 - 0.004 e^{i\delta} & 0.704 \\
0.035 - 0.071 e^{i\delta} & -0.706 - 0.004 e^{i\delta} & 0.704 \\
\end{array} \right) \,,
\label{eq:U2}\\
U({\rm LOW})
&=& \left( \begin{array}{ccc}
0.807 & 0.582 & 0.101 e^{-i\delta} \\
-0.413 - 0.058 e^{i\delta} & 0.574 - 0.042 e^{i\delta} & 0.704 \\
0.413 - 0.058 e^{i\delta} & -0.574 - 0.042 e^{i\delta} & 0.704 \\
\end{array} \right) \,,
\label{eq:U3}\\
U({\rm BIMAX})
&=& \left( \begin{array}{ccc}
{1\over\sqrt2} & {1\over\sqrt2} & 0 \\
-{1\over2} & {1\over2} & {1\over\sqrt2} \\
{1\over2} & -{1\over2} & {1\over\sqrt2} \\
\end{array} \right) \,.
\label{eq:U4}
\end{eqnarray}
Since $\sin^22\theta_{13}$ is not well-known, sometimes we will
consider the LAM solution with $\sin^22\theta_{13} = 0.004$, for which
\begin{equation}
U({\rm LAM^\prime})
= \left( \begin{array}{ccc}
0.850 & 0.525 & 0.032 e^{-i\delta} \\
-0.372 - 0.019 e^{i\delta} & 0.602 - 0.012 e^{i\delta} & 0.707 \\
0.372 - 0.019 e^{i\delta} & -0.602 - 0.012 e^{i\delta} & 0.707 \\
\end{array} \right) \,.
\label{eq:U5}
\end{equation}

Scenarios 1--3 in Table~\ref{tab:scenarios} represent three-neutrino
oscillation explanations of the atmospheric and solar deficits, with the
LAM, SAM, and LOW solar options, respectively. We do not address the VO
solar solution separately, since the sub-leading $\delta m^2$ effects
will not be significant and the LOW and VO scenarios will be
indistinguishable. Scenario~4 with bimaximal atmospheric and LAM
mixing~\cite{bimax} is interesting because the leading oscillation
decouples in the $\nu_e\to\nu_\mu$ channel ($U_{e3} = 0$) and the
sub-leading oscillations will therefore be more visible. However, in this
scenario there are no matter effects on $\nu_e$ propagation and the sign
of $\delta m^2$ cannot be measured; also CP will be conserved.

The size of the Jarlskog invariant (modulo $\sin\delta$) is also shown in
Table~\ref{tab:scenarios}; it is largest for the LAM and LOW scenarios
(which have only one small angle), smaller for SAM (which has two small
angles), and vanishes for the BIMAX scenario (in which one angle is
zero). Since the observability of $CP$ violation depends on both $J$
{\it and} the size of the sub-leading oscillation scale $\delta m^2_{21}$
(see Eq.~(\ref{eq:deltaP2})), one expects appreciable $CP$ violation only
in the LAM scenario.

\begin{figure}
\centering\leavevmode
\epsfxsize=5in\epsffile{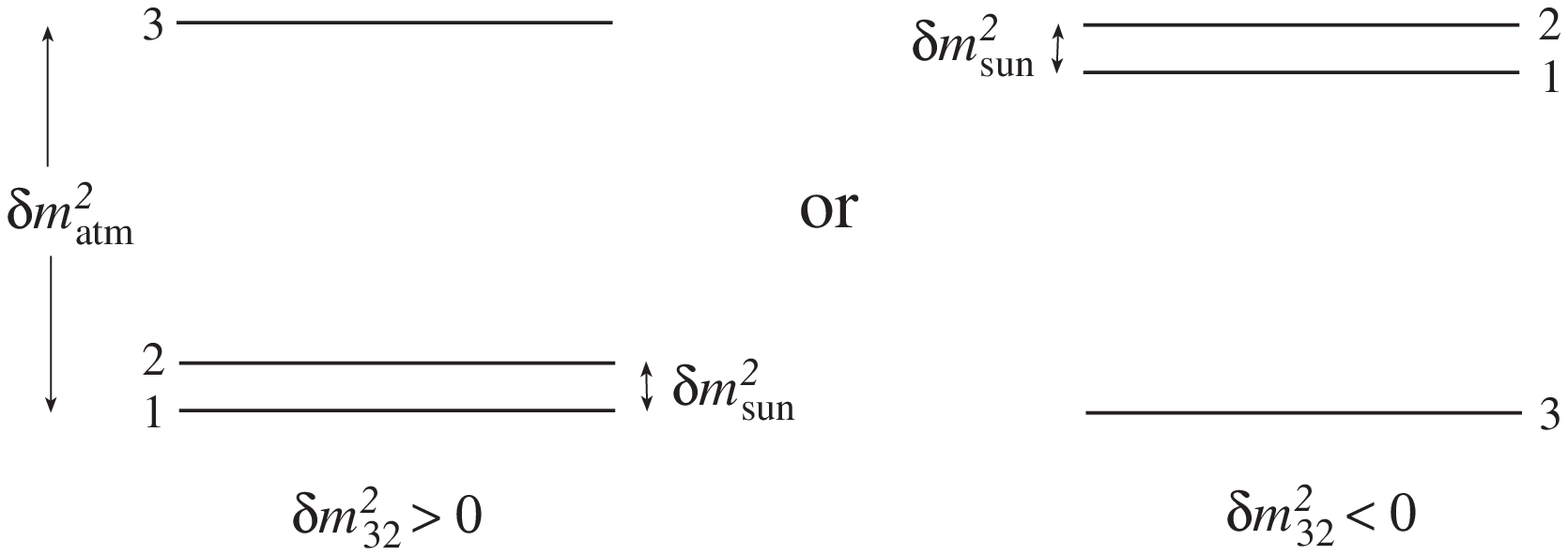}
\bigskip
\caption[]{Two patterns of three-neutrino mass spectra that can explain the
atmospheric and solar neutrino anomalies: (a)~$\delta m_{\rm 32}^2 > 0$,
(b)~$\delta m_{\rm 32}^2 < 0$.}
\label{fig:mass-order}
\end{figure}

The neutrino mass ordering can in principle be determined by the effects
of matter on the leading electron neutrino oscillation\cite{BGRW}. To
illustrate this, Fig.~\ref{fig:mass-order} shows the two possible
three-neutrino mass patterns. Figure~\ref{fig:mass-order}a, with one
large mass $m_3$, has atmospheric neutrino oscillations with $\delta
m_{32}^2 >0$. Figure~\ref{fig:mass-order}b, with two large masses $m_2,
m_1$, has $\delta m_{32}^2 <0$. For scenario~4, with no $\nu_e$
participation in the leading oscillations, there are no matter effects
at the $\delta m_{32}^2$ scale to determine the sign of $\delta m^2_{32}$.

\section{An Entry-Level Neutrino Factory}

Neutrino factories require the development of new accelerator
sub-systems which are technically challenging. The R\&D required for a
full-intensity muon source might take many years.  It is
reasonable to consider a strategy in which the R\&D needed for the first
neutrino factory is minimized by building, as a first step, a
muon source that provides just enough muon decays per year to make
contact with the interesting physics.  If we also wish to minimize the
cost of an entry-level facility, we must minimize the muon
acceleration system and hence the energy of the muons decaying within
the storage ring.  In this section we consider, within the framework of
the scenarios listed in Table~I, the minimum muon energy and beam
intensity needed at an ENuF.

\begin{figure}
\epsfxsize4.0in
\centerline{\epsffile{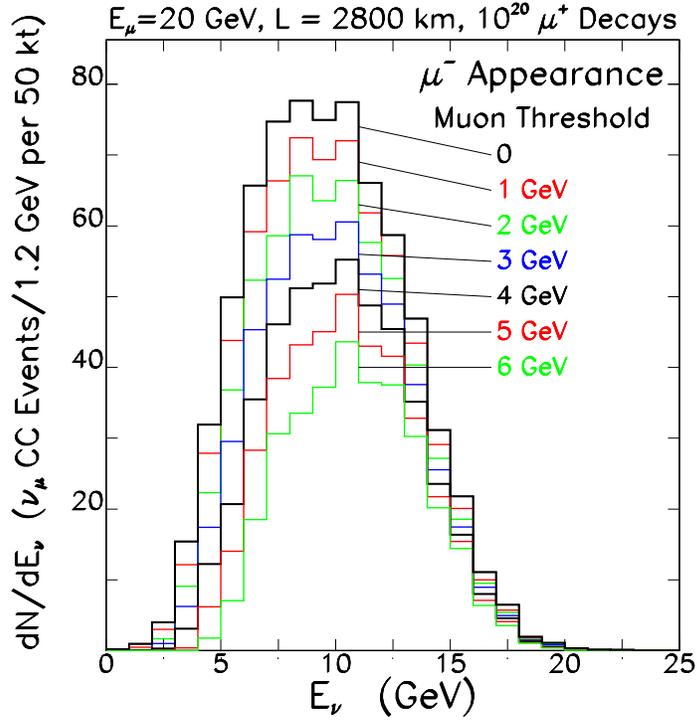}}
\bigskip
\bigskip
\caption[]{
Predicted measured energy distributions (including the detector
resolution  function) for $\nu_e \rightarrow \nu_\mu$  CC events tagged
by wrong-sign muons when 20~GeV positive muons are stored in the
neutrino factory and a 50~kt detector is at $L = 2800$~km. The
distributions are shown for a variety of muon detection threshold
energies $E_{\rm min}$. The oscillation parameters are those 
for the LAM scenario (Table~I).}
\label{fig:threshold}
\end{figure}

We begin by defining our entry-level physics goal. We take this to be
the first observation of $\nu_e\to\nu_\mu$ oscillations at the 10 event
per year level. The signal will be the appearance of CC
interactions which are tagged by a wrong-sign muon. 
To identify signal events 
we must be able to identify muons and measure their charge in the
presence of the accompanying hadronic shower from the remnants of the
target nucleon. Muons can only be cleanly identified and 
measured if their energy exceeds a threshold $E_{\rm min}$, which in 
practice is expected~\cite{det_talks} to be a few GeV. This 
places an effective lower bound on the acceptable energy of the muon
storage ring. To illustrate this, consider the $\nu_e\to\nu_\mu$ signal in a
detector that is 2800~km downstream of a 20~GeV neutrino factory. The
predicted measured energy distributions for CC events tagged by
wrong-sign muons are shown for the LAM scenario (Table~I) in
Fig.~\ref{fig:threshold} as
a function of $E_{\rm min}$.  As $E_{\rm min}$ increases the signal
efficiency decreases.  For example, with $E_{\rm min} =$~ 2 (4) [6]~GeV
the resulting signal loss is 18\% (36\%) [55\%].  In addition, as
$E_{\rm min}$ increases the measured signal distributions become
increasingly biased towards higher energies, and the information on the 
oscillations encoded in the energy distribution is lost. We conclude
from Fig.~\ref{fig:threshold} that with a 20~GeV storage ring we can
probably tolerate an
$E_{\rm min}$ of a few GeV, but would not want to decrease the storage
ring energy below 20~GeV. Hence, we will adopt 20~GeV as the minimum
storage ring energy for an ENuF. 
In the following, our calculations include a muon threshold 
$E_{\rm min} = 4$~GeV, and for simplicity we assume the detection 
efficiency is 0 for signal events with $E_\mu < E_{\rm min}$ and 
1 for $E_\mu > E_{\rm min}$.

\begin{figure}[t]
\epsfxsize3.4in
\centerline{\epsffile{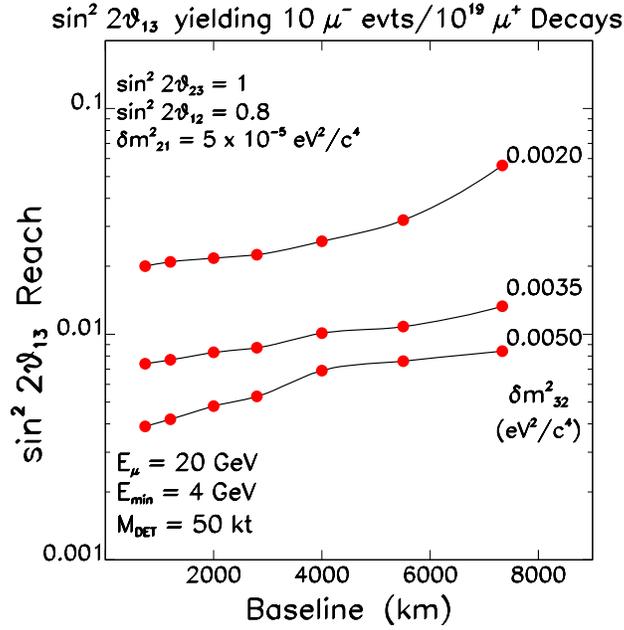}}
\medskip
\caption[]{Reach in $\sin^22\theta_{13}$ for the observation of 
10 $\mu^-$ events from $\nu_e \rightarrow \nu_\mu$ oscillations, 
shown versus baseline for three values of $\delta m^2_{32}$ 
spanning the favored SuperK range. The other oscillation parameters 
correspond to the LAM scenario in Table~I. 
The curves correspond to 
$10^{19} \mu^+$ decays in a 20~GeV neutrino factory with 
a 50~kt detector, and a minimum muon detection threshold of 4~GeV.}
\label{fig:minampvsL1}
\end{figure}

We next consider the muon beam intensity required to meet our
entry-level physics goal if the storage ring energy is 20~GeV.  To
minimize the required beam intensity we must maximize the detector mass
$(M)$. Recently 50~kt has been considered~\cite{det_talks} as a plausible
although ambitious $M$. We therefore choose $M = 50$~kt.  In addition to
the neutrino factory beam properties and the detector mass, the signal
event rate will depend upon the baseline and the oscillation parameters.
Since, to a good approximation, the signal rate is proportional to 
$\sin^22\theta_{13}$, 
it is useful to define the $\sin^22\theta_{13}$ ``reach" for an
experiment as the value of $\sin^22\theta_{13}$ for which our physics
goal (in this case the observation of 10 signal events per year) will be
met. Setting the number of useful muon decays per year to $10^{19}$, the
$\sin^22\theta_{13}$ reach is shown in Fig.~\ref{fig:minampvsL1} as a
function of baseline and $\delta m_{32}^2$ with the other oscillation 
parameters corresponding to the LAM scenario.  The calculational methods
are described in Ref.~\cite{BGW,BGRW}.  The $\sin^22\theta_{13}$ reach
degrades slowly as $L$ increases,  and improves with increasing
$|\delta m_{32}^2|$, varying by about a factor of 5 over the
$\delta m_{32}^2$ range currently favored by the SuperK results. Note
that, for $\delta m_{32}^2$ in the center of the SuperK range, our entry
level goal would be met with a 20~GeV storage ring and $10^{19}$ decays
per year provided $\sin^22\theta_{13}$ exceeds approximately 0.01, which
is more than an order of magnitude below the currently excluded region.

We now consider how the muon beam intensity required to meet our entry-level 
physics goal varies with the storage ring energy.
We will choose a baseline of 2800~km, motivated by a consideration of
the physics sensitivity of an upgraded ENuF 
(see next section). The number of muon decays required to
meet our goal is shown in Fig.~\ref{fig:minfluxvsE} versus
the muon storage ring energy for the LAM, SAM, LOW, and BIMAX
oscillation scenarios.  Note that: 
\begin{description}
\item{(i)} The energy dependent intensities needed for the SAM and LOW 
scenarios are indistinguishable. 
\item{(ii)} Due to the contributions from sub-leading
oscillations, the intensity needed for
the LAM scenario is slightly less than needed for the SAM and LOW
scenarios. 
\item{(iii)} With a 20~GeV storage ring $2 \times 10^{18}$ muon decays 
per year would meet our entry level physics goals for the LAM, SAM, 
and LOW scenarios. The dependence of the required muon intensity $I$ 
on the storage ring energy is approximately given by $I \propto E^{-1.6}$. 
\item{(iv)} For the BIMAX scenario in which $\sin^22\theta_{13} = 0$ only 
the sub-leading $\delta m^2$ scale contributes to the 
signal. With a 20~GeV storage ring a few 
$\times 10^{20}$ muon decays per year would be needed to observe 
$\nu_e\to\nu_\mu$ oscillations. 
Although this scenario would be bad news for a low-intensity neutrino 
factory, oscillations driven by the sub-leading $\delta m^2$ scale might be studied with a higher intensity muon source. 
\end{description}
It is straightforward to use the curves in Fig.~\ref{fig:minfluxvsE}
to infer the  intensity required to meet our entry level goals for 
LAM, SAM, and LOW-type scenarios with values of 
$\sin^22\theta_{13}$ other than 0.04. 
For example, if $\sin^22\theta_{13} = 0.01$ 
(a factor of 20 below the currently 
excluded value, and a factor of 4 below the value used for the 
curves in Fig.~\ref{fig:minfluxvsE}) we must multiply the 
beam intensity indicated in Fig.~\ref{fig:minfluxvsE} by a factor of 4 to
achieve our  entry-level goal. 
Provided a 50~kt detector with good signal efficiency is practical, 
within the framework of LAM, SAM and LOW-type scenarios, 
a 20~GeV storage ring in which there are
$10^{19}$ muon decays per year in the beam forming straight section
would enable the first observation of $\nu_e\to\nu_\mu$ oscillations 
provided $\sin^22\theta_{13} > 0.01$. 
Note that if $\sin^2 2\theta_{13} > 0.01$ the next generation
of long baseline accelerator experiments (e.g.\ MINOS) are
expected make a first observation of $\nu_\mu \rightarrow \nu_e$ 
oscillations. These experiments would not be able to measure
matter effects and determine the sign of $\delta m^2_{32}$.
Furthermore, if $\sin^2 2\theta_{13} \ll 0.01$ then with several  
years of running the entry level neutrino factory data could
be used to place a limit on this parameter about an order of
magnitude below the MINOS limit.

Consider next the prospects for exploiting $\nu_e\to\nu_\mu$ 
measurements to determine the sign of $\delta m_{32}^2$. 
In ref.~\cite{BGRW} we have shown in the LAM
scenario that the sign of $\delta m_{32}^2$ can be determined by comparing
the wrong-sign muon rates and/or the associated CC event energy
distributions when respectively positive and negative muons are
stored in the ring. The most sensitive technique to discriminate 
$\delta m_{32}^2 > 0$ from $\delta m_{32}^2 < 0$ 
would be to take data when there were alternately positive and 
negative muons stored in the neutrino factory, and measure 
the resulting wrong-sign muon event energy distributions 
together with the $\nu_\mu\to\nu_\mu$ and 
$\overline{\nu}_\mu\to\overline{\nu}_\mu$ event energy distributions.
The four distributions can then be 
simultaneously fit with the oscillation parameters $\delta m_{32}^2$ 
(including 
its sign), $\sin^22\theta_{13}$, and $\sin^22\theta_{23}$ left as 
free parameters~\cite{next_paper}. In the following we take a simpler 
approach to demonstrate that, provided $L$ is large enough, a 
neutrino factory that permitted the observation of 10~$\nu_e\to\nu_\mu$ 
events per year would also enable the sign of $\delta m_{32}^2$ 
to be determined. 

\begin{figure}
\centering\leavevmode
\epsfxsize=4in\epsffile{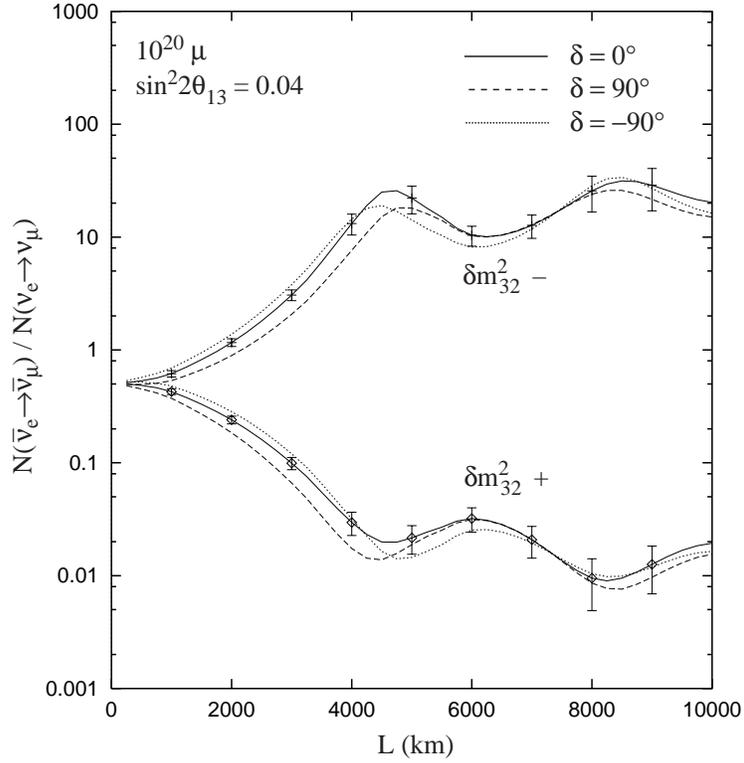}
\bigskip
\caption[]{The ratio $R$ of
$\bar\nu_e \to \bar\nu_\mu$ to $\nu_e \to \nu_\mu$  event
rates at a 20~GeV neutrino factory
for $\delta = 0$ and $\pm\pi/2$. The upper group of curves
is for $\delta m^2_{32} < 0$, the lower group is for
$\delta m^2_{32} > 0$, and the statistical errors correspond to
$10^{20}$ muon decays of each sign and a 50~kt detector.
The oscillation parameters correspond to the LAM solar solution 
with $|\delta m^2_{32}| = 3.5\times10^{-3}$~eV$^2$ and 
$\sin^22\theta_{13}=0.04$. 
}
\label{fig:r1}
\end{figure}

We begin by defining the ratio: 
\begin{equation}
R_{e\mu} = {N(\bar\nu_e\to\bar\nu_\mu) \over
N(\nu_e\to\nu_\mu)  }
\end{equation}
$R_{e\mu}$ is just the ratio of wrong-sign muon rates 
when respectively negative and positive muons are stored 
in the neutrino factory. 
Figure~\ref{fig:r1} shows $R_{e\mu}$ as a function of $L$ for 
$E_\mu = 20$~GeV and 
$\delta m_{32}^2 = \pm 3.5 \times 10^{-3}$~eV$^2$/c$^4$. 
Note that when $L > 2000$~km the ratio $R_{e\mu}$ for positive 
$\delta m_{32}^2$ is more than a factor of 5 greater than 
the value for negative $\delta m_{32}^2$. 
As an example, consider a 50~kt detector 2800~km downstream of a 
20~GeV storage ring in which there are
$10^{19}$ muon decays per year in the beam forming straight section. 
Suppose that $\sin^22\theta_{13} = 0.01$, and assume we know 
that $|\delta m_{32}^2| \sim 3.5 \times 10^{-3}$~eV$^2$/c$^4$ 
from $\nu_\mu\to\nu_\mu$ measurements, for example. 
If we store positive muons in the neutrino factory after one 
year we would expect to observe 11 wrong-sign muon events 
if $\delta m_{32}^2 > 0$ but only 2 events if $\delta m_{32}^2 < 0$. 
To reduce the uncertainties due to the lack of precise knowledge 
of the other oscillation parameters, we can then take data with negative 
muons stored. We would then expect to observe less than $2$~wrong-sign muon 
events per year if $\delta m_{32}^2 > 0$, but 6~events per year if 
$\delta m_{32}^2 < 0$. Clearly with these statistics and several years 
of data taking the sign of $\delta m_{32}^2$ could be established. 
From this example we conclude that with a few years of data taking 
a neutrino factory that enabled the observation of 10~$\nu_\mu\to\nu_\mu$ 
events per year would also enable the sign of $\delta m_{32}^2$ to be 
determined provided 
the baseline was sufficiently long ($L > 2000$~km) so that the 
prediction for $R_{e\mu}$ changes by a large factor ($>5$) when 
the assumed sign of $\delta m_{32}^2$ is changed.
Hence, provided $\sin^22\theta_{13} > 0.01$, 
our entry-level neutrino factory would make the first 
observation of $\nu_e\to\nu_\mu$ oscillations, measure 
$\sin^22\theta_{13}$, and determine the pattern of neutrino masses. 

\section{One Step Beyond an Entry Level Neutrino Factory}

An entry-level neutrino factory is attractive if there is a beam
intensity and/or energy upgrade path that enables a more comprehensive
physics program beyond the initial observation of $\nu_e\to\nu_\mu$ 
oscillations and determination of the sign of $\delta m_{32}^2$. 
In this section we consider the energy and/or intensity
upgrades needed to achieve a reasonable upgrade physics goal, 
which we take to be 
the first observation of $\nu_e\to\nu_\tau$ oscillations.

In ref.~\cite{BGW} we have shown that in a 
LAM-like scenario the $\nu_e\to\nu_\tau$ oscillation event 
rates in a multi-kt detector are expected to be significant if 
$E_\mu$ is 20~GeV or greater.
In the following we consider the neutrino factory
beam energy and intensity needed to make a first observation 
of $\nu_e\to\nu_\tau$ oscillations at the 10~event level in 
one year of running with a fully efficient detector (or several 
years with a realistic detector). Our detector must be able to 
measure $\tau$ appearance and, to separate the signal from 
$\nu_\mu\to\nu_\tau$ oscillation backgrounds, measure the 
sign of the charge of the $\tau$. Hybrid emulsion detectors in 
an external magnetic field provide an example of a candidate detector 
technology that might be used. Consideration of detector 
technologies and their performance is under study~\cite{study}, 
and is outside the scope of the present paper. We will take $M = 5$~kt 
as a plausible, but aggressive, choice for the detector mass.

\begin{figure}[t]
\epsfxsize3.5in
\centerline{\epsffile{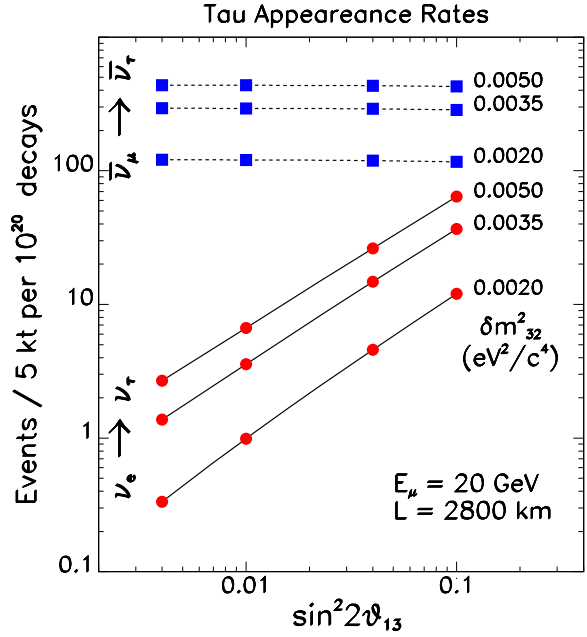}}
\medskip
\caption[]{$\tau$ CC appearance rates in a 5~kt detector 2800~km 
downstream of a 20~GeV neutrino factory in which there are 
$10^{20} \mu^+$ decays in the beam-forming straight section. 
The rates are shown as a function of 
$\sin^22\theta_{13}$ and $\delta m_{32}^2$ with the other oscillation 
parameters corresponding to the LAM scenario in Table~I.
The top 3 curves are the predictions for $\bar\nu_\mu \rightarrow \bar\nu_\tau$ 
events and the lower curves are for $\nu_e \rightarrow \nu_\tau$ events.}
\label{fig:nutaurates}
\end{figure}

Consider first an intensity-upgraded ENuF, namely 
a 20~GeV storage ring in which there are $10^{20}$ muon decays per
year in the beam forming straight section. The $\tau$ appearance 
rates from $\nu_e\to\nu_\tau$ and $\nu_\mu\to\nu_\tau$ oscillations 
are shown as a function of $\sin^22\theta_{13}$ and $\delta m_{32}^2$ 
in Fig.~\ref{fig:nutaurates} for a 5~kt detector at $L = 2800$~km. 
In contrast to the $\nu_\mu\to\nu_\tau$ background rate, which is 
independent of $\sin^22\theta_{13}$, the 
$\nu_e\to\nu_\tau$ signal rate increases linearly with 
$\sin^22\theta_{13}$. Note that for the LAM scenario with 
$\sin^22\theta_{13} = 0.04$ and $\delta m_{32}^2 = 0.0035$~eV$^2$, 
there are about 10 signal events per $10^{20} \mu^+$ decays, and 300 
$\nu_\mu\to\nu_\tau$ background events. Thus it is desirable that 
the $\tau$ sign mis-determination be less than of order 1 in 100. 
Whether this requirement can be met by placing hybrid
emulsion or liquid Argon detectors in a magnetic field, 
or by the development of new $\nu_\tau$ detector 
technology, remains to be seen.

\begin{figure}[h]
\epsfxsize3.5in
\centerline{\epsffile{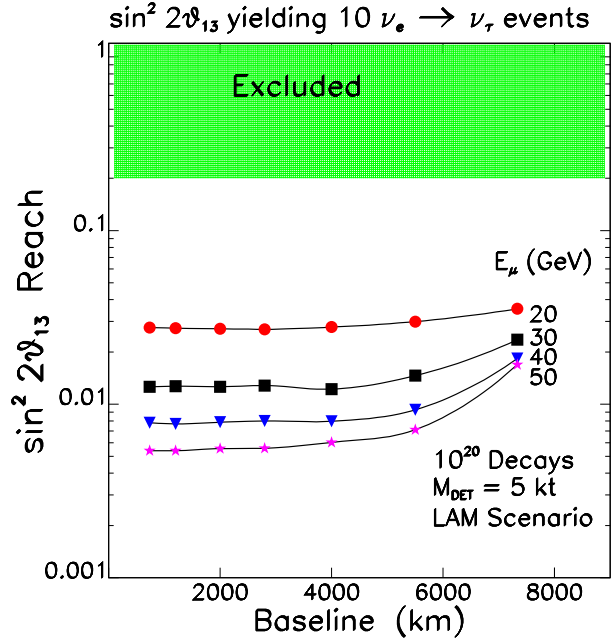}}
\medskip
\caption[]{Reach in $\sin^22\theta_{13}$ for the observation of 
10 events from $\nu_e \rightarrow \nu_\tau$ oscillations, 
shown versus baseline for four storage ring energies. 
The oscillation parameters correspond to the LAM scenario in Table~I. 
The curves correspond to 
$10^{20}$ $\mu^+$ decays in a 20~GeV neutrino factory with 
a 5~kt detector.}
\label{fig:nutaureach}
\end{figure}

Next consider the dependence of the $\sin^22\theta_{13}$ reach 
(for detecting 10 $\nu_e\to\nu_\tau$ events) on 
$L$ and the storage ring energy. Fixing $\delta m_{32}^2 =
0.0035$~eV$^2$, the $\sin^22\theta_{13}$ reach is shown in
Fig.~\ref{fig:nutaureach} to improve with energy, and to be almost independent 
of $L$ over the range considered except at the highest energies 
and longest baselines, for which the reach is degraded. For 
$L \sim 3000$~km, an energy upgrade from 20~GeV to 50~GeV would 
improve the reach by about a factor of 5. 
The energy dependence of the muon intensity required to meet our 
$\nu_e\to\nu_\tau$ discovery goal is summarized in Fig.~\ref{fig:minfluxvsE}. 
We conclude that a neutrino factory consisting of a 
20~GeV storage ring in which there are $10^{20}$ muon decays 
per year in the beam forming straight section would enable the 
first observation of $\nu_e\to\nu_\tau$ oscillations in 
LAM, SAM, and LOW-type scenarios with $\sin^22\theta_{13} > 0.01$
provided a 5~kt detector with good $\tau$ signal efficiency 
and charge-sign determination is practical.

\section{Measuring the CP Non-Conserving Phase}

\begin{figure}
\centering\leavevmode
\epsfxsize=4in\epsffile{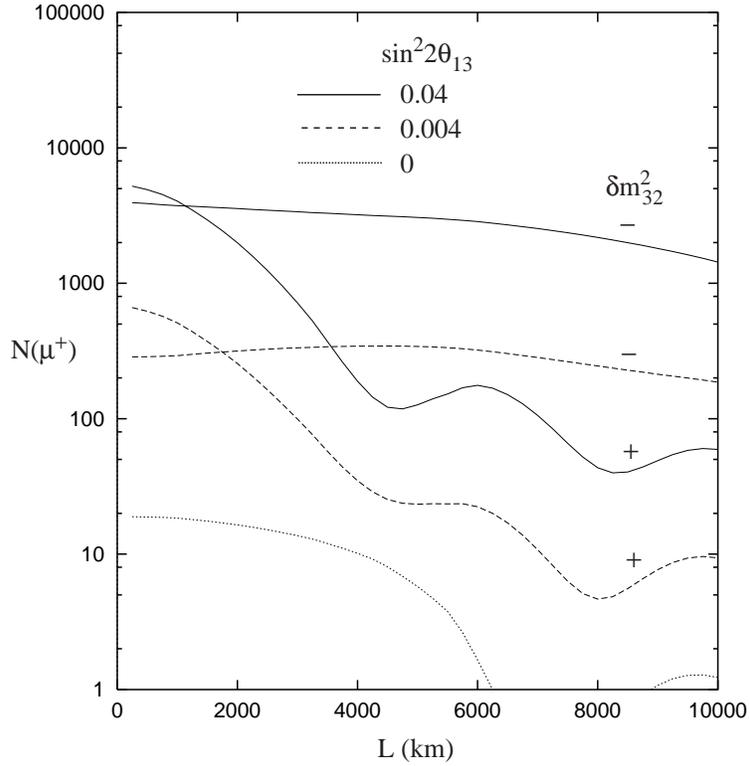}
\bigskip
\caption[]{The number of wrong-sign $\mu^+$ CC events shown versus $L$,
with $10^{21}$ $\mu^-$, a 50~kt detector, and the LAM solar solution
with $|\delta m^2_{32}| = 3.5\times10^{-3}$~eV$^2$ and $\delta = 0$, for
the cases $\sin^22\theta_{13} = 0.04$ (solid curves), $0.004$ (dashed),
and $0$ (dotted). Results for both signs of $\delta m^2_{32}$ are shown.}
\label{fig:Nmu+}
\end{figure}

\begin{figure}
\centering\leavevmode
\epsfxsize=4in\epsffile{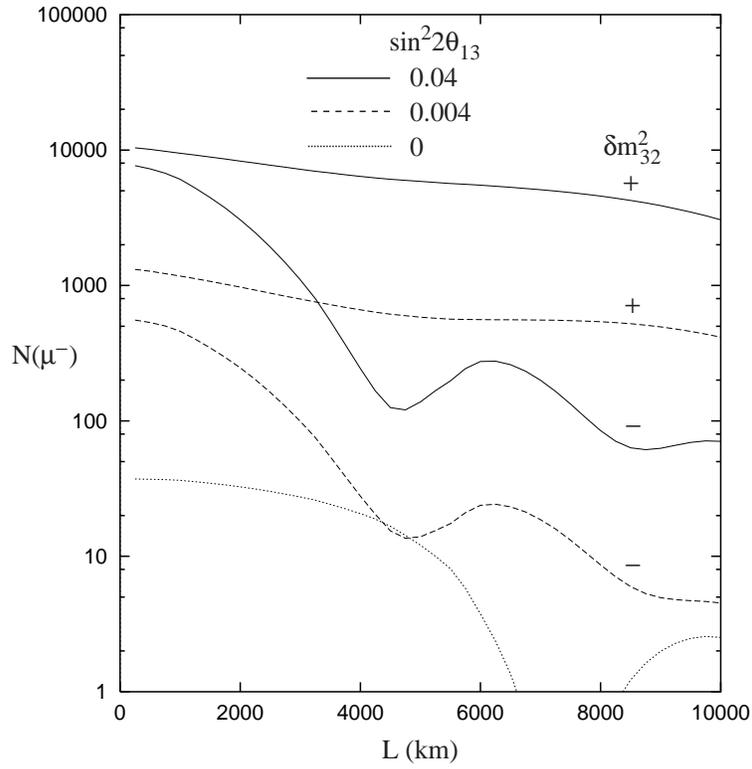}

\caption[]{The same as Fig.~\ref{fig:Nmu+} for wrong sign $\mu^-$ events
from a $\mu^+$ beam.}
\label{fig:Nmu-}
\end{figure}

In the event that the solar solution is LAM, CP non-conserving effects
may be large enough to allow a measurement of the MNS phase $\delta $ at
a high-intensity neutrino factory~\cite{derujula,cervera,freund}.  The total rate of
appearance events very strongly depends on the value of
$\sin^22\theta_{13}$. For example, Figs.~\ref{fig:Nmu+} and
\ref{fig:Nmu-} show the event rates versus $L$ for the LAM solution in
Table~\ref{tab:scenarios} with $\delta = 0$ and $\sin^22\theta_{13} =
0.04$, $0.004$, and $0$. For $L$ less than about 5000~km the event rates
for the BIMAX solution are about 25\% higher than the
$\sin^22\theta_{13} = 0$ curve. Although the rates decrease
significantly with decreasing $\sin^22\theta_{13}$, even for
$\sin^22\theta_{13} = 0$ there is a residual signal from the sub-leading
oscillation in the LAM scenario which may be detectable.

\begin{figure}
\centering\leavevmode
\epsfxsize=4in\epsffile{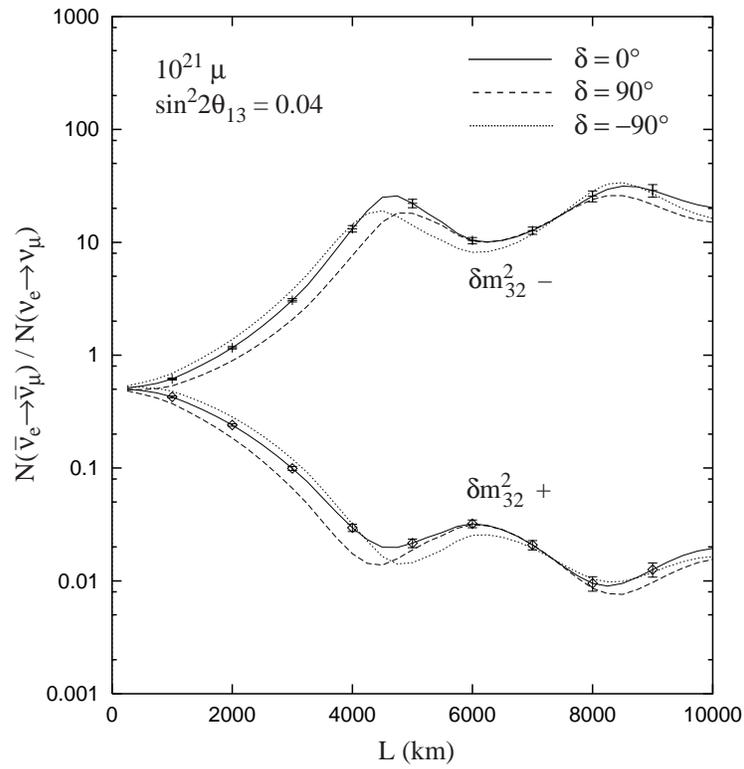}

\caption[]{Same as Fig.~\ref{fig:r1} except for $10^{21}$ muons.}
\label{fig:r2}
\end{figure}

\begin{figure}
\centering\leavevmode
\epsfxsize=4in\epsffile{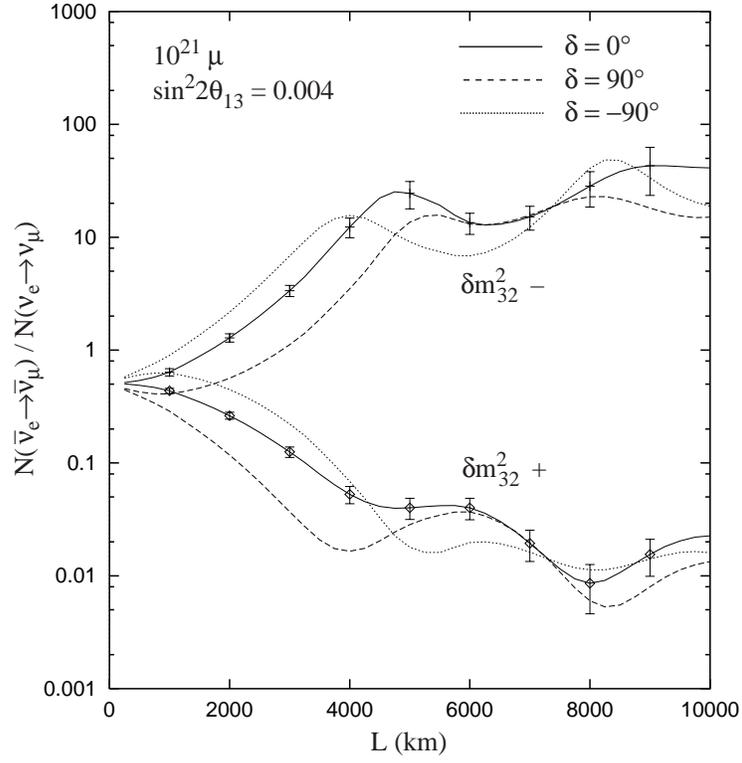}

\caption[]{Same as Fig.~\ref{fig:r1} except for $10^{21}$ muons and
$\sin^22\theta_{13} = 0.004$.}
\label{fig:r3}
\end{figure}

\begin{figure}
\centering\leavevmode
\epsfxsize=4in\epsffile{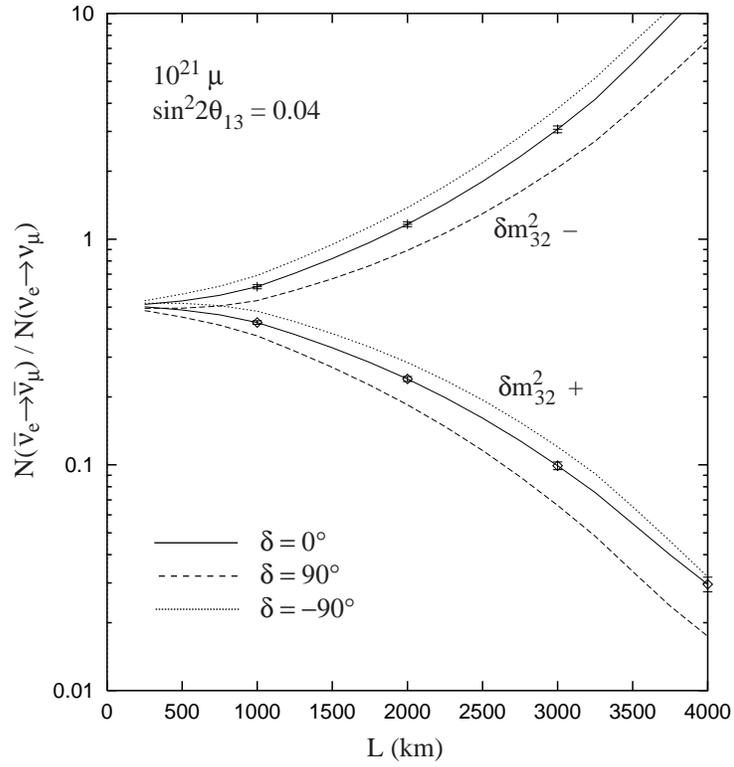}

\caption[]{Same as Fig.~\ref{fig:r2} for the range $0 \le L \le
4000$~km.}
\label{fig:r4}
\end{figure}

Figures~\ref{fig:r1}, \ref{fig:r2}, and \ref{fig:r3} show predictions for the CP dependent
ratio $R_{e\mu}$ 
versus the baseline $L$ for a 50~kt detector in the LAM scenario with
$\delta m^2_{32} = 3.5\times10^{-3}$~eV$^2$. The error bars are
representative statistical uncertainties. These
figures present the results of calculations with decays/year and
$\sin^22\theta_{13}$ values of ($10^{20}$, 0.04), ($10^{21}$, 0.04), and
($10^{21}$, 0.004), respectively. Results for phases $\delta = 0^\circ$,
$\delta = 90^\circ$, and $\delta = -90^\circ$ are shown in each case,
for both positive and negative values of $\delta m_{32}^2$. For these
values of $\sin^22\theta_{13}$ the event rates show a strong dependence
on the sign of $\delta m^2_{32}$
(due to different matter effects for neutrinos and antineutrinos), and a
smaller dependence on the $CP$-violating phase. Figure~\ref{fig:r4} shows the
results for $10^{21}$ muons and $\sin^22\theta_{13} = 0.04$ in finer
detail. Note that at small distances
$R_{e\mu}$ is not unity for $\delta = 0$ (even though matter effects are
small) because the $\bar\nu_\mu$ and $\nu_\mu$ CC cross sections are
different. We note that the $CP$-violating effect is largest in the
range $L \simeq 2000$--3000~km, vanishes for $L \simeq 7000$~km, and
is nonzero but with large uncertainties for $L > 7000$~km.

Similar calculations show that for the SAM and LOW model
parameters $R_{e\mu}$ is essentially independent of $\delta$, verifying
the conclusions of Sec.~III that $CP$ violation is negligible in these
scenarios. The effect of matter in the SAM and LOW scenarios, which
depends on the sign of $\delta m^2_{32}$ and the size of
$\sin^22\theta_{13}$, is similar to the LAM case.

\begin{figure}
\centering\leavevmode
\epsfxsize=4in\epsffile{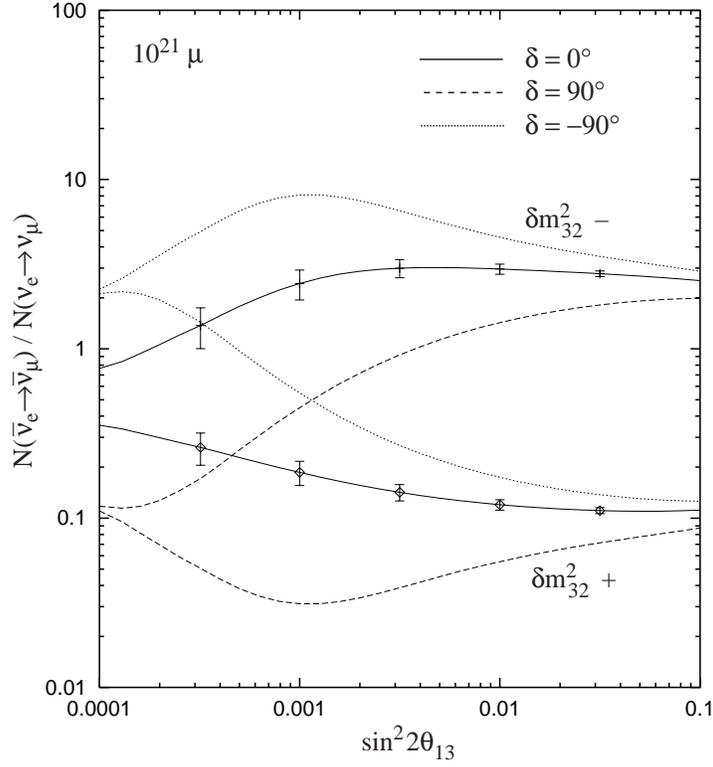}

\caption[]{The ratio of $\bar\nu_e \to \bar\nu_\mu$ to
$\nu_e \to \nu_\mu$ CC events versus $\sin^22\theta_{13}$, with
$L = 2900$~km, $E_\mu = 20$~GeV, $10^{21}$ muons, and a 50~kt detector.
The other oscillation parameters are the same as the LAM scenario in
Table~\ref{tab:scenarios}, and
results for both positive and negative $\delta m^2_{32}$ are shown.
Predictions for maximal $CP$ phases $\delta = 90^\circ$ (dashed curves)
and $\delta =-90^\circ$ (dotted) are compared with the $CP$-conserving
case $\delta =0^\circ$ (solid). The error bars show typical statistical
uncertainties on the measurements.}
\label{fig:rvsamp}
\end{figure}

A nonzero $\sin^22\theta_{13}$ is needed both to determine the sign of
$\delta m^2_{32}$ via matter effects, and to have observable $CP$
violation from the sub-leading scale; however, whether matter or $CP$
violation gives the largest effect depends on the size of
$\sin^22\theta_{13}$. This is illustrated in Fig.~\ref{fig:rvsamp},
which shows the ratio of wrong sign muon events versus
$\sin^22\theta_{13}$ for our representative LAM solar solution with
$10^{21}$ muons/year. For larger values of $\sin^22\theta_{13}$, say
above about 0.001, the sign of $\delta m^2_{32}$ (through matter
effects) has the largest effect on the ratio, while for
$\sin^22\theta_{13} < 0.001$ the value of $\delta$ (which largely
determines the amount of $CP$ violation) has the largest effect.

\begin{figure}
\epsfxsize4.0in
\centerline{\epsffile{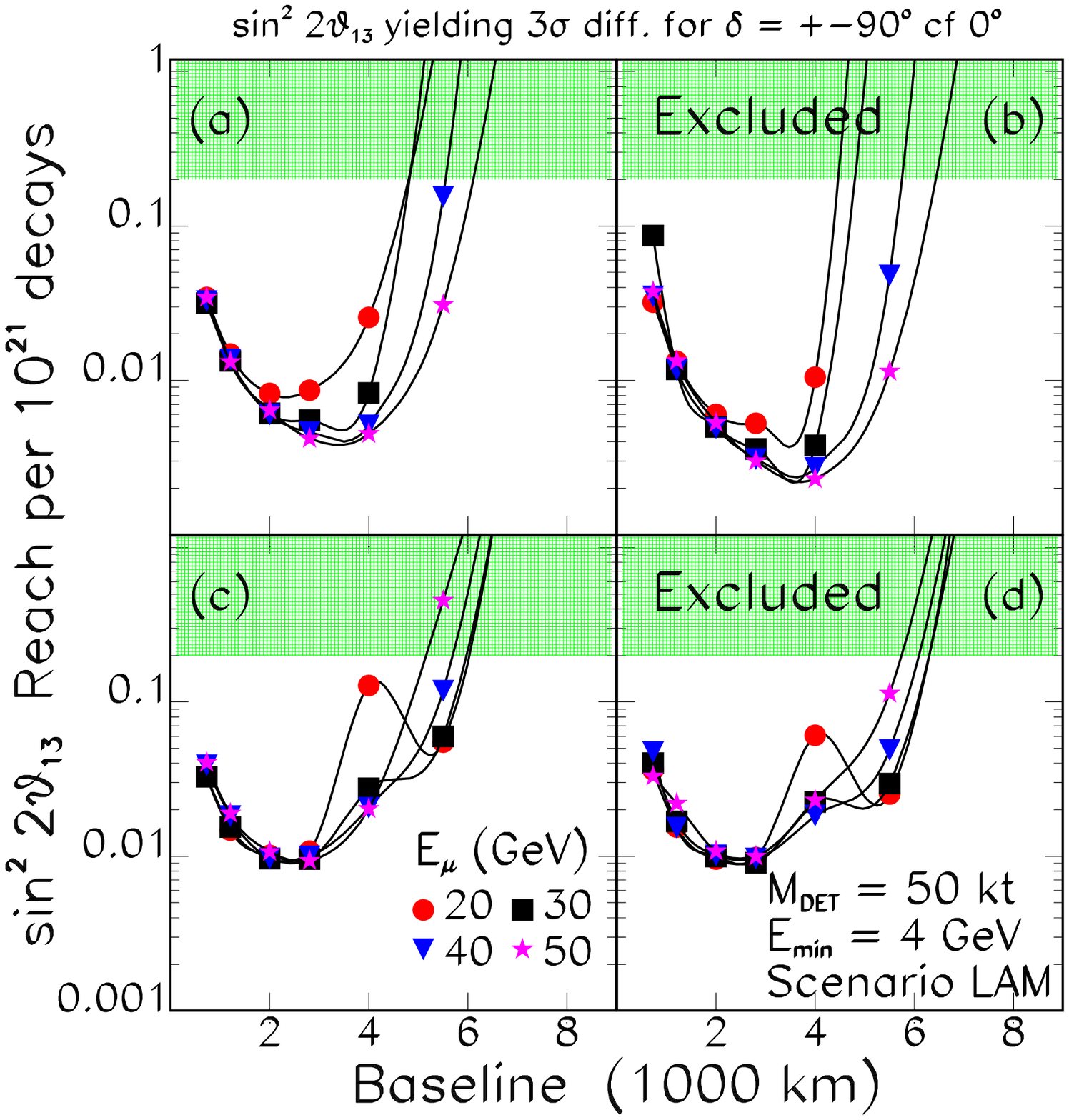}}
\bigskip
\caption[]{Reach in $\sin^22\theta_{13}$ that yields a $3\sigma$
discrimination between
(a) $\delta = 0$ and $\pi/2$ with $\delta m^2_{32} > 0$,
(b) $\delta = 0$ and $\pi/2$ with $\delta m^2_{32} < 0$,
(c) $\delta = 0$ and $-\pi/2$ with $\delta m^2_{32} > 0$, and
(d) $\delta = 0$ and $-\pi/2$ with $\delta m^2_{32} < 0$.
The discrimination is based on a comparison of wrong-sign
muon CC event rates in a 50~kt detector when $10^{21}$ positive and
negative muons alternately decay in the neutrino factory.
The reach is shown versus baseline for four storage ring energies.
The oscillation parameters correspond to the LAM scenario.
}
\label{fig:cpreach1}
\end{figure}

Now consider the neutrino factory energy and intensity needed 
to begin to probe the CP phase $\delta$ in the LAM scenario. 
We will define the 
$\sin^22\theta_{13}$ reach as that value of 
$\sin^22\theta_{13}$ that (with a 50~kt detector and two years of data taking) 
will enable a 3$\sigma$ discrimination between 
(a) $\delta = 0$ and $\delta = \pi/2$ and 
(b) $\delta = 0$ and $\delta = -\pi/2$. The measurement will be based on a
comparison of wrong-sign muon rates when respectively positive and
negative muons are alternately stored in the ring. 
The $\sin^22\theta_{13}$ reach when there are $10^{21}$ decays  
is shown for the 3$\sigma$ discrimination between $\delta = 0$ and 
$\pm\pi/2$ in Fig.~\ref{fig:cpreach1} 
as a function of baseline and stored muon energy. 
The optimum baseline is about 3000~km, for which 
the $\sin^22\theta_{13}$ reach is a little better (worse) than 0.01 
for the $\delta = \pi/2$ ($\delta = -\pi/2$) discrimination, 
and is almost independent of muon energy 
over the range considered (Fig.~\ref{fig:minfluxvsE}). 
Thus, a high intensity 20~GeV neutrino factory 
providing O($10^{21}$) muon decays might begin to probe CP violation 
in the lepton sector if the LAM scenario is the correct description 
of neutrino oscillations, and a 50~kt detector with good 
signal efficiency is practical. This conclusion is consistent with 
results presented in Ref.~\cite{cervera} in which global fits to the 
measured oscillation distributions have been studied to determine 
the precision with which $\delta$ and $\sin^2 2\theta_{13}$ can 
be simultaneously measured at a neutrino factory.

\section{Summary}

We briefly summarize the results of our study of the physics goals of an
entry-level neutrino factory as follows:

\renewcommand{\theenumi}{\roman{enumi}}
\renewcommand{\labelenumi}{(\theenumi)}
\begin{enumerate}

\item An entry-level machine would make 
a first observation of $\nu_e\to\nu_\mu$
oscillations, measure the corresponding amplitude $\sin^22\theta_{13}$,
and determine the sign of $\delta m_{32}^2$.

\item The $\sin^22\theta_{13}$ reach for the first observation of
$\nu_e\to\nu_\mu$ oscillations and the measurement of the sign of $\delta
m_{32}^2$ is insensitive to the solar neutrino oscillation solution (LAM, SAM,
or LOW) within a 3-neutrino framework.

\item A 20~GeV neutrino factory providing $10^{19}$ muon decays per year 
would enable our entry-level physics goals to be met 
provided $\sin^22\theta_{13}>0.01$ and 
a detector with good muon charge-sign determination and a 
mass of 50~kt is practical. 
The required beam intensity might be 
a factor of 2--3 higher or lower depending on
where within the SuperK range the $\delta m_{32}^2$ parameter sits. The event
rates also depend on the muon energy detection threshold ($E_{\rm min} =
2$--5~GeV) and we win or lose a factor of 2 in rates depending on how low
this threshold can be pushed. To determine the sign of $\delta m_{32}^2$ 
a long baseline ($L > 2000$~km) must be chosen.

\item A candidate for an intensity-upgraded neutrino factory would 
be a 20~GeV facility providing $10^{20}$ muon decays per year. 
In addition to the precise determination of the oscillation 
parameters, the upgraded neutrino source would enable the 
first observation of $\nu_e\to\nu_\tau$ oscillations provided 
$\sin^22\theta_{13} > 0.01$ and 
a 5~kt detector with good $\tau$ signal efficiency 
and charge-sign determination is practical.

\item With a high-intensity neutrino factory providing 
a few $\times 10^{20}$ muon decays per year, 
the ratio of $\mu^+/\mu^-$ wrong-sign muon rates might enable 
detection of a maximal CP phase in the case of the LAM
solar solution. If $\sin^22\theta_{13}$ is vanishingly small, 
(for example the bimaximal mixing scenario) a first observation 
of $\nu_e\to\nu_\mu$ oscillations at a high intensity neutrino 
factory might provide a direct measurement of oscillations 
driven by the sub-leading $\delta m^2$ scale, although 
wrong-sign muon backgrounds might be problematic.

\end{enumerate}

\begin{figure}[h]
\epsfxsize3.4in
\centerline{\epsffile{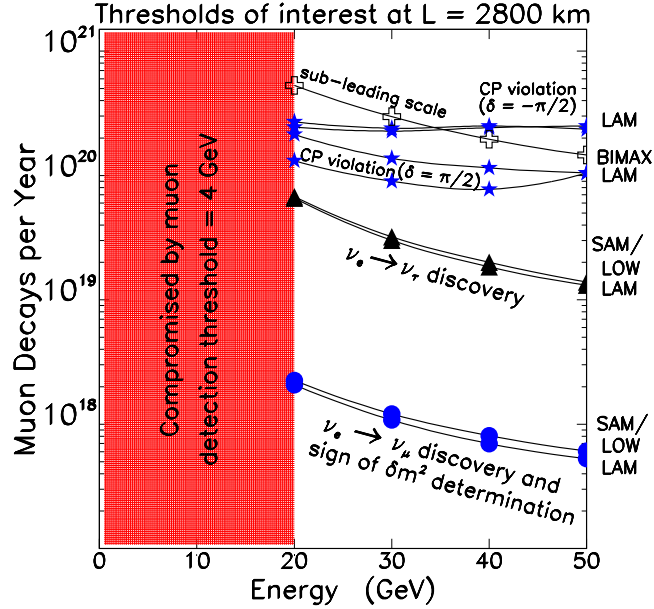}}
\medskip
\caption[]{The required number of muon decays needed in the beam-forming 
straight section of a neutrino factory to achieve the physics goals described 
in the text, shown as a function of storage ring energy for the scenarios 
listed in Table~I. The baseline is taken to be 2800~km, and 
the detector is assumed to be a 50~kt wrong-sign muon 
appearance device with a muon detection threshold of 4~GeV or, for 
$\nu_e \rightarrow \nu_\tau$ appearance, a 5~kt detector.}
\label{fig:minfluxvsE}
\end{figure}

In conclusion, the required number of muon decays per year to achieve 
the various
physics goals of interest are summarized in Fig.~\ref{fig:minfluxvsE}. 
At a 20~GeV neutrino 
factory $\sim10^{19}$ muon decays are required for a thorough search
for $\nu_e \rightarrow \nu_\mu$ appearance, $\sim 10^{20}$ decays 
to search for $\nu_e \rightarrow \nu_\tau$ oscillations, and 
$\sim 10^{21}$ decays  to probe the sub-leading oscillation 
scale and detect CP violation effects in a three-neutrino LAM 
scenario.

\acknowledgments
This research was supported in part by the U.S.~Department of Energy under
Grant No.~DE-FG02-95ER40896 and in part by the University of Wisconsin Research
Committee with funds granted by the Wisconsin Alumni Research Foundation.

\newpage

\renewcommand{\topfraction}{1.0}   
\renewcommand{\bottomfraction}{1.0}
\renewcommand{\textfraction}{0.0}  

\newpage

\newpage

\newpage

\newpage

\begin{thebibliography}{99}   \addtolength{\itemsep}{-.5mm}
\bibitem{atmos1}
Super-Kamiokande Collaboration,
Y. Fukuda et al., Phys. Lett. {\bf B433}, 9 (1998);
Phys. Lett. {\bf B436}, 33 (1998);
Phys. Rev. Lett. {\bf 81}, 1562 (1998);
Phys. Rev. Lett. {\bf 82}, 2644 (1999).

\bibitem{atmos2}
Kamiokande collaboration, K.S. Hirata {\it et al.}, Phys. Lett.
{\bf B280}, 146 (1992); Y. Fukuda {\it et al.}, Phys. Lett.
{\bf B335}, 237 (1994);
IMB collaboration, R. Becker-Szendy {\it et al.}, Nucl. Phys. Proc.
Suppl. {\bf 38B}, 331 (1995);
Soudan-2 collaboration, W.W.M. Allison {\it et al.}, Phys. Lett. {\bf
B391}, 491 (1997);
MACRO collaboration, M. Ambrosio {\it et al.},
Phys. Lett. {\bf B434}, 451 (1998).

\bibitem{kajita99}
T. Kajita, talk presented at the {\it 7th International Symposium on Particles,
Strings and Cosmology (PASCOS\,99)}, Granlibakken, CA, Dec.~1999,\\
http://pc90.ucdavis.edu/talks/plenary/Kajita/

\bibitem{solar1}
B.T. Cleveland {\it et al.}, Nucl. Phys. B (Proc. Suppl.) {\bf 38}, 47
(1995);
GALLEX collaboration, W. Hampel {\it et al.}, Phys. Lett. {\bf B388},
384 (1996);
SAGE collaboration, J.N. Abdurashitov {\it et al.}, Phys. Rev. Lett.
{\bf 77}, 4708 (1996);
Kamiokande collaboration, Y. Fukuda {\it et al.}, Phys. Rev. Lett,
{\bf 77}, 1683 (1996);
Super-Kamiokande collaboration, Y. Fukuda {\it et al.},
Phys. Rev. Lett. {\bf 82}, 2430 (1999);
Phys. Rev. Lett. {\bf 82}, 1810 (1999).

\bibitem{solar2}
J.N. Bahcall, S. Basu, and M.H. Pinsonneault, Phys. Lett.
{\bf B 433}, 1 (1998), and references therein.

\bibitem{solardata}
A. de Gouvea, A. Friedland, and H. Murayama, hep-ph/0002064;
G.L. Fogli, E. Lisi, D. Mintanino, and A. Palazzo, hep-ph/9912231.
M.C.~Gonzalez-Garcia, P.C.~de Holanda, C.~Pe\~na-Garay, and J.W.F.~Valle,
hep-ph/9906469;
J.N. Bahcall, P. Krastev, and A.Yu. Smirnov,
hep-ph/9905220;
V.~Barger and K.~Whisnant, Phys. Lett. {\bf 456}, 54 (1999);
V.~Barger and K.~Whisnant, Phys. Rev. {\bf D59}, 093007 (1999);
J.N. Bahcall, P. Krastev, and A.Yu. Smirnov,
Phys. Rev. {\bf D58}, 096016 (1998);
R. Barbieri, L.J. Hall, D. Smith, A. Strumia, and N. Weiner,
JHEP 9812, 017 (1998);
N.~Hata and P.~Langacker, Phys. Rev. {\bf D56}, 6107 (1997).


\bibitem{wolf-etal}
L. Wolfenstein, Phys. Rev. {\bf D17}, 2369 (1978).

\bibitem{bppw80}
V.~Barger, S.~Pakvasa, R.J.N.~Phillips, and K.~Whisnant, Phys.
Rev. {\bf D22}, 2718 (1980).

\bibitem{langacker}
P. Langacker, J.P. Leveille, and J. Sheiman,
Phys. Rev. {\bf D 27}, 1228 (1983);


\bibitem{mikheyev}
S.P. Mikheyev and A. Smirnov, Yad. Fiz. {\bf 42}, 1441 (1985)
[Sov. J. Nucl. Phys. 42, 913 (1986)].

\bibitem{parke-etal}
S.~Parke, Phys. Rev. Lett. {\bf 57}l, 1275 (1986);
S.P.~Rosen and J.M.~Gelb, Phys. Rev. {\bf D34}, 969 (1986);
W.~Haxton, Phys. Rev. Lett. {\bf 57}, 1271 (1986).

\bibitem{bpw81}
V. Barger, K. Whisnant, R.J.N. Phillips,
Phys. Rev. {\bf D24}, 538 (1981);
S.L.~Glashow and L.M.~Krauss, Phys. Lett. {\bf B190}, 199 (1987).

\bibitem{vlw}
V. Barger and K.~Whisnant, Phys. Lett. {\bf B456}, 54 (1999);
V. Barger, R.J.N. Phillips, and K. Whisnant,
Phys. Rev. Lett. {\bf 65}, 3084 (1990);
{\bf 69}, 3135 (1992);
P.I. Krastev and S.T. Petcov, Phys. Lett. {\bf B 285}, 85 (1992);
Nucl. Phys. {\bf B449}, 605 (1995);
S.L.~Glashow, P.J.~Kernan, and L.M.~Krauss, Phys. Lett. {\bf B445}, 412 (1999).

\bibitem{sno} G.T.~Ewans et al., Physics in Canada {\bf 48}, 2 (1992); see also
the SNO web pages at http://www.sno.phy.queensu.ca/

\bibitem{lsnd}
C. Athanassopoulos et al. (LSND Collab.),
Phys. Rev. Lett. {\bf 77}, 3082 (1996);
Phys. Rev. Lett. {\bf 81}, 1774 (1998).

\bibitem{miniboone}
E. Church et al. (BooNE Collab.), ``A letter of intent
for an experiment to measure $\nu_\mu \rightarrow \nu_e$
oscillations and $\nu_\mu$  at the Fermilab Booster",
May 16, 1997, unpublished.

\bibitem{k2k}
K. Nishikawa et al. (KEK-PS E362 Collab.),
``Proposal for a Long Baseline Neutrino Oscillation Experiment,
using KEK-PS and Super-Kamiokande", 1995, unpublished;
INS-924, April 1992, submitted to J.~Phys. Soc. Jap.;
Y. Oyama, Proc. of the YITP Workshop on Flavor
Physics, Kyoto, Japan, 1998, hep-ex/9803014.

\bibitem{minos}
MINOS Collaboration, ``Neutrino Oscillation Physics at Fermilab: The
NuMI-MINOS Project,'' NuMI-L-375, May 1998.

\bibitem{icanoe} See the ICANOE web page at http://pcnometh4.cern.ch/

\bibitem{opera} See the OPERA web page at http://www.cern.ch/opera/

\bibitem{thesis}D. A. Petyt,``A study of parameter measurement in a
long-baseline
neutrino oscillation experiment", Thesis submitted to Univ. of Oxford,
England, 1998.

\bibitem{geerconf}
S. Geer,
``Neutrino beams from muon storage rings: characteristics
      and physics potential", FERMILAB-PUB-97-389, 1997, presented at the
Workshop on Physics at the First Muon Collider and Front-End
      of a Muon Collider, November, 1997.

\bibitem{geer}
S. Geer, Phys. Rev. {\bf D57}, 6989 (1998).

\bibitem{mucoll}
C. Ankenbrandt et al. (Muon Collider Collaboration),  Phys. Rev. ST Accel.
Beams 2, 081001 (1999).

\bibitem{derujula}
A. De Rujula, M.B. Gavela, and P. Hernandez,
Nucl. Phys. {\bf B547}, 21 (1999).

\bibitem{camp}
M. Campanelli, A. Bueno, and A. Rubbia, hep-ph/9905240.

\bibitem{nu-factory}
R.B.~Palmer et al. (Muon Collider Collababoration),
\newline
http://pubweb/bnl.gov/people/palmer/nu/params.ps

\bibitem{BGW} V. Barger, S. Geer, and K. Whisnant,
Phys.\ Rev. {\bf D 61}, 053004 (2000).

\bibitem{BGRW}V. Barger, S. Geer, R. Raja, and K. Whisnant, Phys. Rev. {\bf D62}, 013004 (2000).

\bibitem{cervera}
A. Cervera, A. Donini, M.B. Gavela, J.J. Gomez Cadenas, P. Hernandez,
O. Mena, and S. Rigolin, hep-ph/0002108.

\bibitem{shrock}
I. Mocioiu and R. Shrock, hep-ph/9910554, hep-ph/0002149.



\bibitem{cabbibo}
N. Cabbibo, Phys. Lett. {\bf B72}, 333 (1978).

\bibitem{bpw80}
V. Barger, K. Whisnant, R.J.N. Phillips,
Phys. Rev. Lett. {\bf 45}, 2084 (1980)

\bibitem{pakvasa}
S.~Pakvasa, in {\it Proc. of the XXth International Conference on High Energy
Physics}, ed. by L.~Durand and L.G.~Pondrom, AIP Conf. Proc. No.~68 (AIP, New
York, 1981), Vol.~2, p.~1164.

\bibitem{barger-CP}
V. Barger, Y.-B. Dai, K. Whisnant, and B.-L. Young,
Phys. Rev. {\bf D59}, 113010 (1999).

\bibitem{freund}
M.~Freund, M.~Lindner, S.T.~Petcov, and A.~Romanino, hep-ph/9912457.


\bibitem{CP}
For other recent discussions of $CP$ violation effects in neutrino
oscillations, see 
H.~Fritzsch and Z.-Z.~Xing, Phys. Rev. {\bf D61}, 073016 (2000); Z.-Z.~Xing, hep-ph/0002246;
D.J. Wagner and T.J. Weiler, Phys. Rev. {\bf D59}, 113007
(1999);
A.M.~Gago, V.~Pleitez, R.Z.~Funchal, hep-ph/9810505;
K.R.~Schubert, hep-ph/9902215;
K.~Dick, M.~Freund, M.~Lindner and A.~Romanino, hep-ph/9903308;
J.~Bernabeu, hep-ph/9904474;
S.M.~Bilenky, C.~Giunti, and W.~Grimus,
Phys. Rev. {\bf D58}, 033001 (1998);
M.~Tanimoto, Prog. Theor. Phys. {\bf 97}, 901 (1997);
J.~Arafune, J.~Sato, Phys. Rev. {\bf D55}, 1653 (1997);  
T.~Fukuyama, K.~Matasuda, H.~Nishiura,
Phys. Rev. {\bf D57}, 5844 (1998);
M.~Koike and J.~Sato, hep-ph/9707203,
Proc.\ of the 5th KEK Meeting on CP Violation and its Origin, Tsukuba,
Japan, March 1997, ed.\ by K. Hagiwara (KEK Proceedings 97-12);
H.~Minakata and H.~Nunokawa, Phys. Lett. {\bf B413}, 369 (1997);
H.~Minakata and H.~Nunokawa, Phys. Rev. {\bf D57}, 4403 (1998);
J.~Arafune, M.~Koike and J.~Sato, Phys. Rev. {\bf D56}, 3093 (1997);
M.~Tanimoto, Phys. Lett. {\bf B435}, 373 (1998).

\bibitem{bernstein}
R. Bernstein and S. Parke, Phys. Rev. {\bf D44}, 2069 (1991).

\bibitem{pantaleone}
J. Pantaleone, Phys. Rev. {\bf D49}, 2152 (1994);
Phys. Rev. Lett. {\bf 81}, 5060 (1998);

\bibitem{lipari}
P. Lipari, hep-ph/9903481;
E. Akhmedov, P. Lipari, and M. Lusignoli,
Phys. Lett. {\bf B300}, 128 (1993);
P. Lipari and M. Lusignoli,
Phys. Rev. {\bf D58}, 073005 (1998);
E. Akhmedov, A. Dighe, P. Lipari, and A.Yu. Smirnov,
Nucl. Phys. {\bf B542}, 3 (1999);
E.K. Akhmedov, Nucl. Phys. {\bf B538}, 25 (1998);
S.T. Petcov, Phys. Lett. {\bf B434}, 321 (1998);
M.V. Chizhov, M. Maris, and S.T. Petcov, hep-ph/9810501;
M.V. Chizhov and S.T. Petcov, hep-ph/9903424;
S.T.~Petcov, hep-ph/9910335;
H.W. Zaglauer and K.H. Schwarzer, Z. Phys. {\bf C40}, 273 (1988);
Q. Liu, S. Mikheyev, and A.Yu. Smirnov,
Phys. Lett. {\bf B440}, 319 (1998);
P.I. Krastev, Nuovo Cimento A {\bf 103}, 361 (1990).
J. Pruet and G.M. Fuller, astro-ph/9904023.
J.~Arafune, J.~Sato, Phys. Rev. {\bf D55}, 1653 (1997);
J.~Arafune, M.~Koike and J.~Sato, Phys. Rev. {\bf D56}, 3093 (1997);
M.~Tanimoto, Prog. Theor. Phys. {\bf 97}, 901 (1997);
H.~Minakata and H.~Nunokawa, Phys. Lett. {\bf B413}, 369 (1997);
H.~Minakata and H.~Nunokawa, Phys. Rev. {\bf D57}, 4403 (1998);
M.~Koike and J.~Sato, hep-ph/9909469;
M.~Koike and J.~Sato, hep-ph/9911258;
T.~Ohlsson and H.~Snellman, J. Math. Phys. {\bf41}, 2768 (2000);
A.~Romanino, hep-ph/9909425.

\bibitem{PREM}
Parameters of the Preliminary Reference Earth Model are given
by A. Dziewonski, Earth Structure,
Global, in ``The Encyclopedia of Solid Earth Geophysics", ed. by D.E. James,
(Van Nostrand Reinhold, New York, 1989) p.~331; also see
R. Gandhi, C. Quigg, M. Hall Reno, and I. Sarcevic,
Astroparticle Physics {\bf 5}, 81 (1996).

\bibitem{mns}
Z. Maki, M. Nakagawa, and S. Sakata,
Prog. Theor. Phys. {\bf 28}, 870 (1962).

\bibitem{keung}
W.-Y. Keung and L.-L. Chau, Phys. Rev. Lett. {\bf 53}, 1802 (1984).

\bibitem{jarlskog}
C. Jarlskog, Z. Phys. {\bf C 29}, 491 (1985); Phys. Rev. {\bf D 35},
1685 (1987).

\bibitem{fermi-lb} See the Fermilab Long-Baseline Workshop web site at\\
http://www.fnal.gov/projects/muon\_collider/nu/study/study.html


\bibitem{bimax}
V. Barger, S. Pakvasa, T.J. Weiler, and K. Whisnant,
Phys. Lett. {\bf B437} 107 (1998);
A.J. Baltz, A.S. Goldhaber, and M. Goldhaber,
Phys. Rev. Lett. {\bf 81}, 5730 (1998).

\bibitem{det_talks}
Talks by P.~Spentzouris, D. Harris, K. McFarland, D. Casper, M. Campanelli, 
and S. Rigolin 
at the two day neutrino factory physics study meeting, Fermilab, 
17-18~February, 2000. 
See http://www.fnal.gov/projects/muon\_collider/nu/study/study.html

\bibitem{next_paper}
V. Barger, S. Geer, R. Raja, and K. Whisnant, Phys. Lett. {\bf B485}, 379 (2000) [hep-ph/0004208].

\bibitem{study}
See http://www.fnal.gov/projects/muon\_collider/nu/study/study.html

\end{thebibliography}
\end{document}